\documentclass[final,3p,times]{elsarticle}
\usepackage{amssymb}
\usepackage{amsmath}
\usepackage{color, textcomp, makecell}
\usepackage[colorlinks=true, allcolors=blue]{hyperref}

\journal{}

\begin{document}

\begin{frontmatter}

\title{Toward Sustainable Polymer Design: A Molecular Dynamics-Informed Machine Learning Approach for Vitrimers}

\author[UW]{Yiwen Zheng}
\author[UW]{Agni K. Biswal}
\author[TUD]{Yaqi Guo}
\author[TUD]{Prakash Thakolkaran}
\author[UW]{Yash Kokane}
\author[AFRL]{Vikas Varshney}
\author[TUD]{Siddhant Kumar}
\author[UW]{Aniruddh Vashisth\corref{cor1}}

\cortext[cor1]{Email: vashisth@uw.edu}

\affiliation[UW]{organization={Department of Mechanical Engineering, University of Washington}, 
            city={Seattle},
            state={WA},
            country={USA}}
\affiliation[TUD]{organization={Department of Materials Science and Engineering, Delft University of Technology}, 
            city={Delft},
            country={The Netherlands}}
\affiliation[AFRL]{organization={Materials and Manufacturing Directorate, Air Force Research Laboratory}, 
            city={Wright-Patterson Air Force Base},
            state={OH},
            country={USA}}
            
\begin{abstract}
Vitrimer is an emerging class of sustainable polymers with self-healing capabilities enabled by dynamic covalent adaptive networks. However, their limited molecular diversity constrains their property space and potential applications. Recent development in machine learning (ML) techniques accelerates polymer design by predicting properties and virtually screening candidates, yet the scarcity of available experimental vitrimer data poses challenges in training ML models. To address this, we leverage molecular dynamics (MD) data generated by our previous work to train and benchmark seven ML models covering six feature representations for glass transition temperature ($T_\mathrm{g}$) prediction. By averaging predicted $T_\mathrm{g}$ from different models, the model ensemble approach outperforms individual models, allowing for accurate and efficient property prediction on unlabeled datasets. Two novel vitrimers are identified and synthesized, exhibiting experimentally validated higher $T_\mathrm{g}$ than existing bifunctional transesterification vitrimers, along with demonstrated healability. This work explores the possibility of using MD data to train ML models in the absence of sufficient experimental data, enabling the discovery of novel, synthesizable polymer chemistries with superior properties. The integrated MD-ML approach offers polymer chemists an efficient tool for designing polymers tailored to diverse applications.
\end{abstract}




\end{frontmatter}

\section{Introduction}

Polymers play a crucial role in a wide array of industries, including coating, microelectronics, automobile, and aerospace. However, their performance decays over time as molecular bonds break under conditions such as mechanical stress and temperature during utilization. In traditional thermosets and thermoplastics polymers, this damage is irreversible since the rupture of covalent bonds cannot be reversed. This leads to the formation of cracks and eventual material failure \cite{young2011introduction}. As a result, most end-of-life polymer products end up as waste after failure, posing significant challenges to sustainability. The inability to repair degraded polymers leads to product replacement and increases economic costs in plastics manufacturing. In addition, the lack of intrinsic ability to repair molecular damage in polymers limits the recyclability of used thermoplastics and thermosets.

\begin{figure}[t]
\centering
\includegraphics[width=\linewidth]{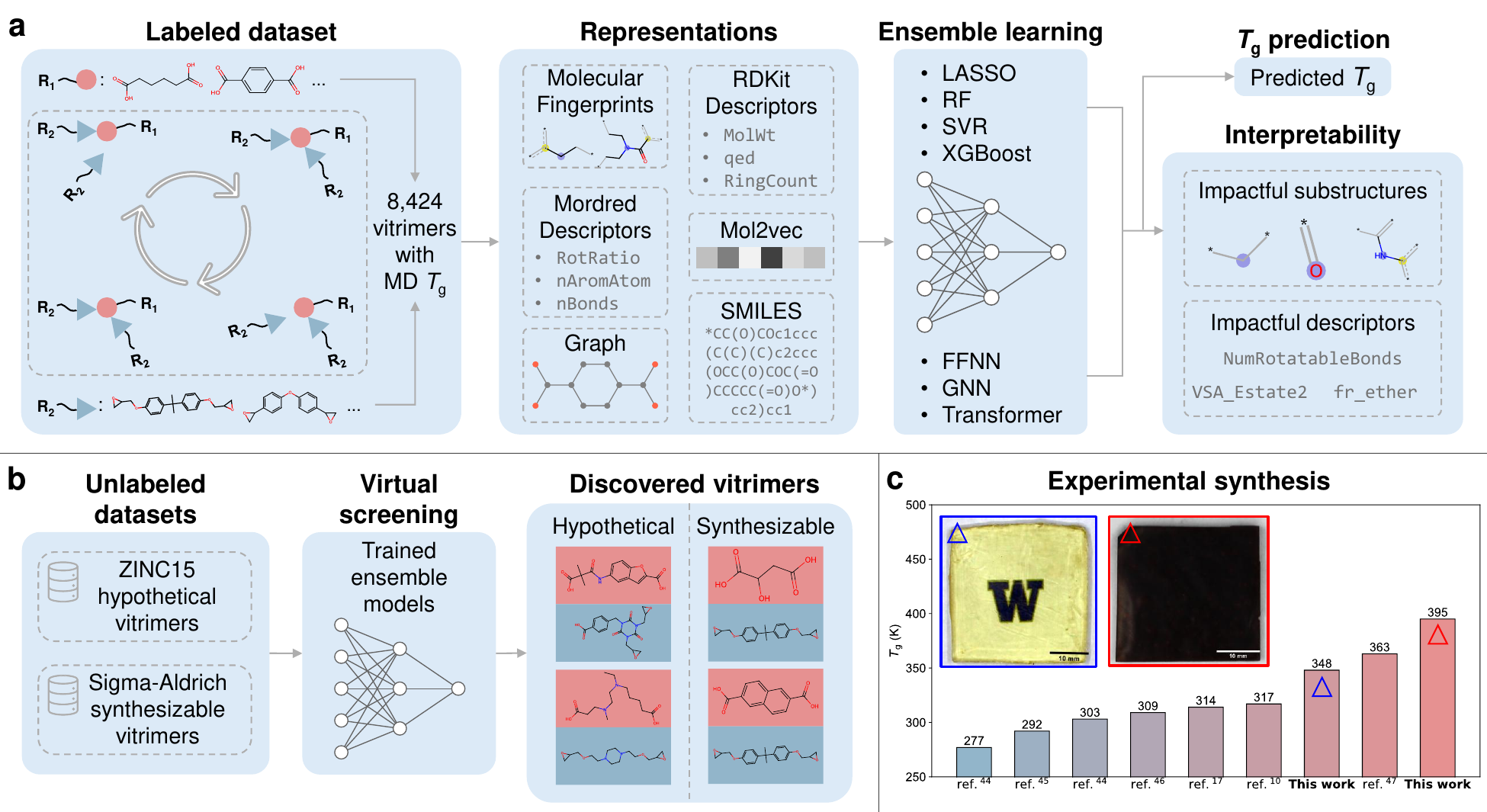}
\caption{Overview of this work. (a) Leveraging a previously developed vitrimer dataset, the performance of seven ML models covering six feature representation types is benchmarked. The ensemble model, which averages predictions from different models, achieves the best performance. We also provide interpretability into the ML models and identify the most impactful descriptors and substructures. (b) The trained ensemble model is applied to predict $T_\mathrm{g}$ of unlabeled datasets and novel vitrimers are discovered. (c) Two screened vitrimers are selected for experimental synthesis and characterization, demonstrating high $T_\mathrm{g}$ compared with bifunctional transesterification vitrimers present in the literature.}
\label{overview}
\end{figure}

Since the introduction by Leibler et al. \cite{montarnal2011silica, capelot2012catalytic, jin2019malleable, krishnakumar2020vitrimers}, a type of healable polymers called vitrimers have gained significant attention as a potential solution to the plastic pollution issue. Combining the reprocessability of thermoplastics and superior mechanical properties of thermosets, vitrimers can potentially reduce plastic production and costs in polymer recycling. Vitrimers are featured by dynamic covalent adaptive networks (CANs) which enable building blocks to attach to and detach from each other via bond exchange reactions, which brings healability and recyclability (Figure \ref{overview}a). Vitrimers can be classified based on the the types of bond exchange reactions including transesterification, disulfide exchange, and imine exchange reactions \cite{jin2019malleable}. However, the thermo-mechanical properties of currently available vitrimers are constrained by the limited variety of commercially available molecular building blocks (i.e., monomers) used in their synthesis. Therefore, to broaden the applications of vitrimers, it is essential to efficiently identify the available or synthesizable monomers that constitute vitrimers with desirable properties.

Traditionally, novel polymer discovery has been accomplished in a trial-and-error fashion \cite{hoogeboom2022equivariant}. For a given set of polymers, chemists characterize the properties of each one of them by experiments and selected those with desired properties. Since experimental synthesis of novel polymer chemistries is costly, molecular dynamics (MD) simulations have been favored as an alternative method to characterize polymers. MD is a simulation technique that bridges quantum mechanics and classical mechanics, and it has been extensively utilized to accelerate the discovery process \cite{hansson2002molecular}. MD simulations have provided valuable insights into how polymer molecular structures influence mechanical properties \cite{vashisth2018accelerated}, glass transition temperature \cite{yu2001polymer}, and self-healing performance \cite{kamble2022reversing}. Despite these advancements, large-scale computational screening using MD or other simulation techniques remains computationally expensive even with the progress in high-performance computing \cite{kranenburg2009challenges}. Consequently, the exploration of polymer design space is typically constrained and fails to cover the vast design space of molecular compositions.

Recent developments in machine learning (ML) methods have opened a new path to accelerate polymer discovery by orders of magnitude. ML models are trained using available data to predict polymer properties (labels) from molecular structures (features). The trained models are employed to predict the properties of unlabeled polymers (whose properties are unknown) and the polymers with promising predicted properties are selected for further validation. This process is also known as virtual screening, which has been applied to design polymers with various properties including glass transition temperature ($T_\mathrm{g}$) \cite{tao2021machine}, thermal conductivity \cite{thakolkaran2024deep}, bandgap \cite{batra2020polymers}, and gas uptake properties \cite{yang2022machine}. Generally, there are three considerations when applying machine learning to polymer property prediction:
\begin{itemize}
    \item Data: High-quality datasets are necessary for training accurate ML models. The largest polymer database PolyInfo \cite{otsuka2011polyinfo} contains $\sim$ 18,000 polymers with experimentally measured properties but lack sufficient data in specific properties. For example, there are only 173 data points of thermal conductivity. As an experimental database, recently synthesized polymers are also absent in PolyInfo. To address these challenges, high throughput MD simulations allow for creating a large and consistent dataset to train ML models. Researchers have established and curated polymer datasets of various properties including glass transition temperature \cite{zheng2025ai}, free volume \cite{tao2023machine}, thermal conductivity, \cite{thakolkaran2024deep} and ionic conductivity \cite{xie2022cloud}. ML models have been trained using these datasets for virtual screening of polymers and novel insights into structure-property relationships have been identified.
    \item Representations: Monomers or polymer repeating units consist of atoms and bonds, which, in the original form, are not interpretable to ML models. To achieve accurate predictions, it is essential to select an appropriate way to convert molecular structures into machine-readable numerical feature representations. Molecular descriptors, which are theoretically derived values incorporating physicochemical information of molecules, serve as a commonly used representation \cite{rdkit, moriwaki2018mordred}. Another prevalent representation is molecular fingerprints, which are vectors with substructure occurrences as entries \cite{rogers2010extended}. Molecular structures can also be described using string notations such as simplified molecular-input line-entry system (SMILES) \cite{weininger1988smiles}. Graph-based representations have drawn increasing attention due to the inherent graph nature of molecules, where atoms or substructures serve as nodes and bonds as edges. The selection of the best representation type is typically case-specific and requires empirical trials.
    \item ML models: The choice of ML models is partially dependent on the selected representation of polymers. For example, graph neural networks (GNNs) are specialized for graphs while Transformers are more suitable for sequence-based input such as SMILES \cite{wu2020comprehensive, vaswani2017attention}. Molecular descriptors and fingerprints are more universal and accommodate both shallow and deep ML models such as linear regression, random forest, and feed-forward neural networks. Similar to representations, the choice of best ML models is also empirical.
\end{itemize}

Despite the growing applications of ML for polymers, efforts to design vitrimers using ML remain limited. Yan et al. \cite{yan2023overcoming} employed a variational autoencoder (VAE)-based virtual screening approach to identify vitrimers with desirable glass transition temperature ($T_\mathrm{g}$) and recyclability. In their method, vitrimers were encoded as latent vectors using a pretrained VAE, while a feedforward neural network (FFNN) predicted $T_\mathrm{g}$ from these latent vectors. However, although the VAE was pretrained on $\sim$ 420,000 drug-like molecules, the property predictor was trained on a small experimental dataset (389 data points for $T_\mathrm{g}$) manually curated from the literature, resulting in limited predictive accuracy. In our previous work \cite{zheng2025ai}, we addressed this limitation by conducting large-scale molecular dynamics (MD) simulations to calculate the $T_\mathrm{g}$ of 8,424 hypothetical vitrimers and calibrating calculated $T_\mathrm{g}$ against available experimental data. This dataset, along with an unlabeled set of $\sim$ 1 million hypothetical vitrimers, was used to train a VAE for the inverse design of vitrimers with targeted $T_\mathrm{g}$. While the generative model successfully learned the distribution of valid vitrimers and generated novel chemistries, it did not explicitly account for synthesizability, making the proposed vitrimers difficult to synthesize. In contrast, the virtual screening approach ensures the synthesizability of designed vitrimers by restricting the search space to a predefined unlabeled dataset consisting of vitrimers derived from commercially available monomers.

In this work, we introduce an integrated MD-ML virtual screening framework to discover bifunctional transesterification vitrimers composed of carboxylic acids and epoxides in an equimolar ratio. Leveraging open source MD-generated data from our previous study \cite{zheng2025ai}, we systematically evaluate the performance of ML models in predicting vitrimer glass transition temperature ($T_\mathrm{g}$). We benchmark seven ML models (linear regression, random forest, support vector regression, gradient boosting, feedforward neural network, graph neural network, and Transformer) across six molecular representations, including molecular fingerprints, RDKit \cite{rdkit} descriptors, Mordred \cite{moriwaki2018mordred} descriptors, Mol2vec embeddings \cite{jaeger2018mol2vec}, SMILES, and graphs. Our findings indicate that an ensemble learning approach, which averages predictions from multiple models, outperforms individual models (Figure \ref{overview}a). Our analysis of feature importance further provides interpretability by identifying key factors that influence the $T_\mathrm{g}$ of vitrimers. Using the trained ensemble model, we screen an unlabeled vitrimer dataset containing approximately 1 million hypothetical vitrimers, identifying candidates with extreme (high and low) $T_\mathrm{g}$ values beyond the training domain which are further validated through MD simulations. To ensure synthesizability, we conduct an additional screening of vitrimers composed of commercially available carboxylic acids and epoxides, selecting promising candidates for experimental evaluation. Two novel vitrimers are synthesized and characterized, demonstrating experimentally higher $T_\mathrm{g}$ than any bifunctional transesterification vitrimer reported in the literature. This study demonstrates the feasibility of using MD-generated data to train ML models in the absence of available experimental data and successfully applies this approach to the design and validation of novel vitrimers with exceptional properties through experimental synthesis and characterization. The proposed framework serves as an effective tool for polymer chemists to design synthesizable polymers with tailored properties for diverse applications.

\section{Methods}

\subsection{Datasets}

\begin{table}[t]
    \centering
    \begin{tabular}{c c c c}
        \Xhline{2\arrayrulewidth}
        \textbf{Datasets} & \textbf{Size} & \textbf{With $T_\mathrm{g}$} & \textbf{Source} \\
        \hline
        Labeled dataset & 8,424 & Yes & ZINC15 \cite{sterling2015zinc} \\
        Hypothetical unlabeled dataset & 991,576 & No & ZINC15 \cite{sterling2015zinc} \\
        Synthesizable unlabeled dataset & 259 & No & Sigma-Aldrich \cite{SigmaAldrich} \\
        \Xhline{2\arrayrulewidth}
    \end{tabular}
    \caption{Three vitrimers datasets used in this work.}
    \label{dataset}
\end{table}

We utilize the vitrimer $T_\mathrm{g}$ dataset curated in our previous work \cite{zheng2025ai}, which comprises 1 million hypothetical vitrimers randomly sampled from 2.5 billion possible combinations of 50,000 carboxylic acid molecules and 50,000 epoxide molecules derived from the ZINC15 database \cite{sterling2015zinc}. Of these, 8,424 vitrimers have $T_\mathrm{g}$ values calculated via molecular dynamics (MD) simulations. A Gaussian process (GP) model is employed to calibrate the MD-calculated $T_\mathrm{g}$ values against experimental data from the literature. For this study, the subset of 8,424 vitrimers with $T_\mathrm{g}$ labels serves as the training dataset for ML models, which is referred to as the labeled dataset. The remaining vitrimers within the 1 million set are treated as the hypothetical unlabeled dataset. Further details on the datasets and MD simulations can be found in Supporting Information.

In this work, we also curate a synthesizable dataset to facilitate the experimental synthesis of screened vitrimers. This dataset is constructed by selecting 37 commercially available bifunctional carboxylic acids and 7 bifunctional epoxides from the Sigma-Aldrich website \cite{SigmaAldrich}, yielding a total of 259 vitrimers that can be synthesized using off-the-shelf chemicals. The sizes and sources of all three datasets are provided in Table \ref{dataset}.

\subsection{Feature engineering}

We consider six different feature representations as inputs for machine learning models: molecular fingerprints, RDKit descriptors \cite{rdkit}, Mordred descriptors \cite{moriwaki2018mordred}, Mol2vec embeddings \cite{jaeger2018mol2vec}, SMILES \cite{weininger1988smiles}, and graph-based representations. For each vitrimer in the dataset, the repeating unit is obtained by opening the epoxide rings and reacting them with carboxyl groups (see Figure S1, Supporting Information). These repeating units are then converted into the specified feature representations, which serve as direct inputs for the ML models. Further details are discussed below.

\begin{itemize}
    \item For molecular fingerprints, the repeating units are transformed into 2048-bit Morgan fingerprints \cite{morgan1965generation, rogers2010extended} with a radius of 3 using the RDKit package \cite{rdkit}. Each bit in the fingerprint vector represents the frequency of a specific chemical substructure within the vitrimer, a method known as frequency-based molecular fingerprints. This approach has been shown to outperform traditional binary fingerprints, where each bit only indicates the presence (1) or absence (0) of a substructure \cite{tao2021benchmarking}. We select 200 substructures that appear most frequently across the labeled dataset. Substructures with zero variance (i.e., those with identical occurrence counts across all labeled vitrimers) are removed. The final fingerprint vector consists of 195 entries, each corresponding to a specific substructure present in the vitrimers.
    \item Molecular descriptors for each vitrimer are computed using the RDKit \cite{rdkit} and Mordred \cite{moriwaki2018mordred} packages. Since certain descriptors are not well defined for some vitrimer chemistries, only those that are valid across both the labeled and hypothetical unlabeled datasets are retained. Additionally, descriptors with zero variance are removed. After filtering, 166 RDKit descriptors and 620 Mordred descriptors are selected.
    \item Mol2vec is an unsupervised learning-based approach that represents molecules as fixed-length numerical vectors by learning embeddings of molecular substructures \cite{jaeger2018mol2vec}. In this work, we utilize a pretrained Mol2vec model to encode vitrimers into 300-dimensional vector representations, capturing their structural and chemical characteristics.
    \item SMILES: each vitrimer repeating unit is converted into canonical SMILES using the RDKit package \cite{rdkit} where the asterisk symbol (*) denotes connection sites for the repeating unit.
    \item Following the approach described by Hickey et al. \cite{hickey2024applying}, vitrimers are represented as molecular graphs, where heavy atoms serve as nodes and bonds as edges. Each node is characterized by atomic features including atom type, number of hydrogen atoms, valency, neighbor degree and aromaticity. Edge features include bond type, as well as indicators for whether the bond is conjugated or part of a ring. An adjacency matrix is also extracted with binary indicators of connectivity between each pair of nodes.
\end{itemize} 

\subsection{Machine learning models}

With vitrimer repeating units encoded into various feature representations, we proceed with training and evaluating different ML models for $T_\mathrm{g}$ prediction. The schematic overview of all ML models is presented in Figure S2 (Supporting Information). We train least absolute shrinkage and selection operator (LASSO), random forest (RF), support vector regression (SVR), extreme gradient boosting (XGBoost), and feedforward neural network (FFNN) models using molecular fingerprints, Mol2vec embeddings, RDKit descriptors, and Mordred descriptors. Graph Neural Networks (GNNs) and Transformer models are trained using graph-based representations and SMILES strings, respectively. A total number of 22 models are trained and benchmarked in this work. A brief introduction to each ML model is provided in Supporting Information. The LASSO, RF, and SVR models are implemented using the scikit-learn library \cite{pedregosa2011scikit}, while the XGBoost model is built using the XGBoost package \cite{chen2016xgboost}. The FFNN, GNN, and Transformer models are implemented in PyTorch \cite{paszke2019pytorch}, with the GNN and Transformer architectures adapted from previous implementations by Hickey et al. \cite{hickey2024applying} and Xu et al. \cite{xu2023transpolymer}, respectively. More details on model architectures, training procedures, and hyperparameter tuning are provided in the Supporting Information. To explore possible enhancement in prediction accuracy, we also employ an ensemble learning approach by averaging the $T_\mathrm{g}$ values predicted by different models. This strategy has been used in previous studies for polymer property prediction \cite{alfaraj2023model, esmaeili2023accelerated, gao2024accurately}. We evaluate various combinations of all 22 ML models and identify the ensemble that yields the highest prediction performance on the test set.

The trained ML models not only enable efficient $T_\mathrm{g}$ prediction but also provide molecular insights into the key factors influencing vitrimer $T_\mathrm{g}$. To quantify feature importance, we perform Shapley additive explanations (SHAP) analysis \cite{lundberg2017unified} on the trained LASSO, RF, XGBoost, and FFNN models using molecular fingerprints, RDKit descriptors, and Mordred descriptors. SVR is excluded from this evaluation due to high computational cost. For molecular fingerprints, RDKit descriptors and Mordred descriptors, we compute the average SHAP values (i.e., feature importance scores) across all considered ML models and training data. The most impactful substructures and descriptors are identified based on the highest absolute average SHAP values, providing interpretability into the factors governing vitrimer $T_\mathrm{g}$.

After training all models and selecting the best-performing one, we proceed with screening the unlabeled datasets. Each vitrimer in these datasets is preprocessed using the previously discussed feature representations, and its $T_\mathrm{g}$ is predicted using the trained model. Promising candidates are identified for further validation. For candidates from the hypothetical dataset, MD simulations are performed to compute their $T_\mathrm{g}$, following the same simulation and calibration protocols used in generating the labeled dataset (details provided in the Supporting Information). For candidates from the synthesizable dataset, two vitrimers are selected for experimental synthesis and characterization based on chemical intuition, synthesizability and cost.

\subsection{Experimental synthesis and characterization}

Two vitrimers are synthesized by crosslinking bisphenol A diglycidyl ether (DGEBA) with D,L-malic acid (MA) and 2,6-naphthalenedicarboxylic acid (NA), both purchased from Sigma-Aldrich \cite{SigmaAldrich}. The crosslinking process involves stirring a mixture of the acids with the epoxide and 5 mol \% of the catalyst triazabicyclodecene (TBD) in the presence of 5 mL of acetone. The mixture is stirred on a hot plate at temperatures ranging from 140 to 170 \textdegree C. Subsequently, the slurry is left in the beaker for 5 to 10 minutes to remove the acetone. The degassed resin is then transferred into a mold and covered with a top mold for further crosslinking process in a hot press. The MA-based resin is cured at 140 \textdegree C for 5 hours, while the NA-based resin was cured at 280 \textdegree C for 6 hours under a constant pressure of 1 MPa. The glass transition temperature ($T_\mathrm{g}$) of the cured samples is determined using differential scanning calorimetry (DSC) at a heating rate of 5 \textdegree C per minute.

\section{Results}

\subsection{Machine learning $T_\mathrm{g}$ prediction}

\begin{figure}[t]
\centering
\includegraphics[width=\linewidth]{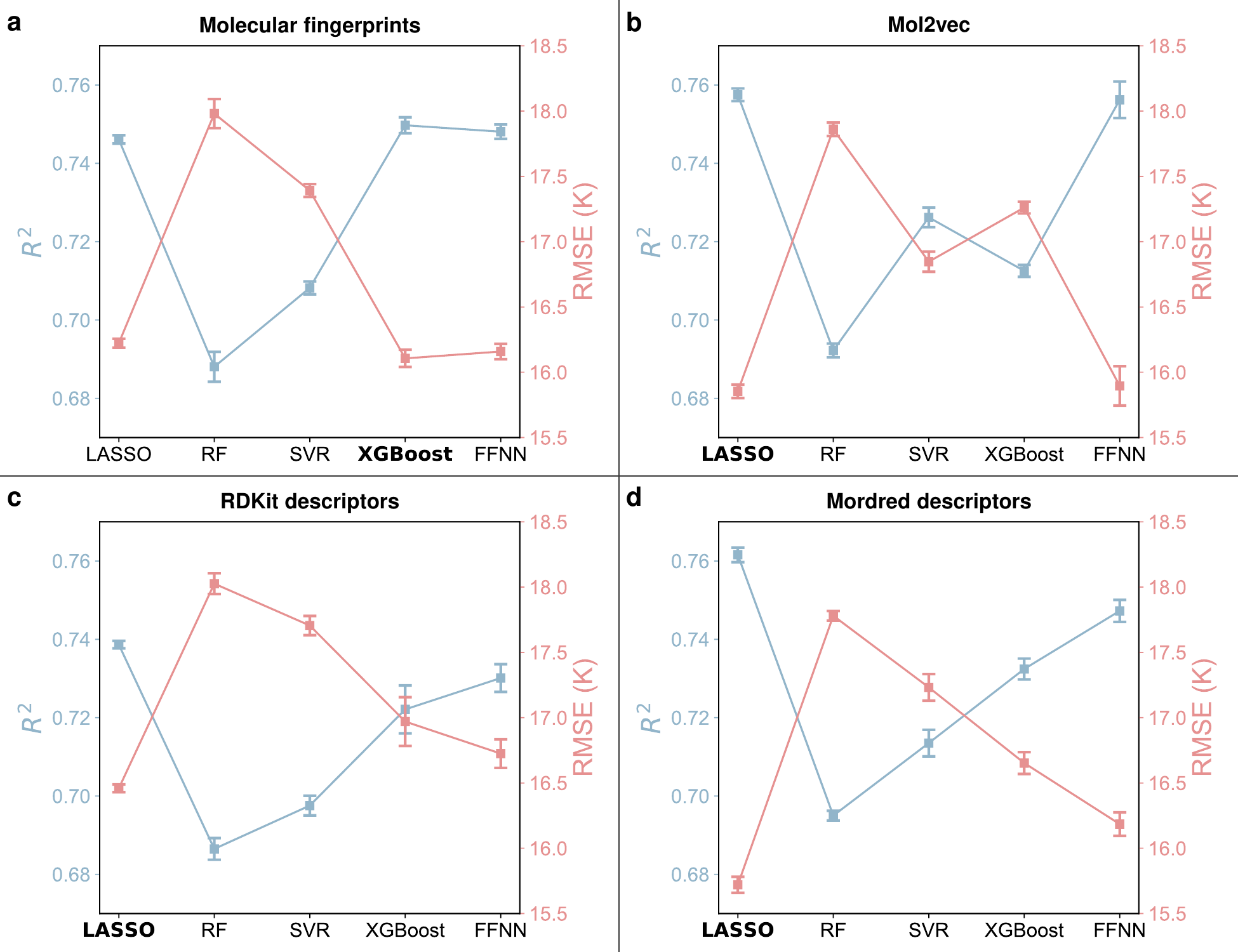}
\caption{Coefficient of determination ($R^2$) and root mean square error (RMSE) between ground truth $T_\mathrm{g}$ and predicted $T_\mathrm{g}$ of LASSO, RF, SVR, XGBoost and FFNN models using (a) molecular fingerprints, (b) Mol2vec embeddings, (c) RDKit descriptors and (d) Mordred descriptors. The error bars indicate standard deviation in the metrics across five-fold cross validation. The best-performing models are highlighted in bold.}
\label{benchmark}
\end{figure}

The performance of five ML models across four feature representations for $T_\mathrm{g}$ prediction on the test set is summarized in Figure \ref{benchmark}. XGBoost achieves the highest accuracy using molecular fingerprints, while LASSO outperforms other models for the remaining three representations. Among 20 (out of overall 22) models evaluated, LASSO with Mordred descriptors (denoted as LASSO\_Mordred) demonstrates the highest predictive accuracy, achieving a coefficient of determination ($R^2$) of 0.76 and a root means square error (RMSE) of 15.72 K. The superior performance of LASSO Mordred can be attributed to the high dimensionality of Mordred descriptors, which include 620 features, a greater number than any other feature representation (195, 300, 166 for molecular fingerprints, Mol2vec embeddings and RDKit descriptors, respectively). Additionally, LASSO incorporates an L1 regularization term, enforcing sparsity in model coefficients, thereby reducing overfitting and improving the accuracy of $T_\mathrm{g}$ predictions.

\begin{figure}[t]
\centering
\includegraphics[width=\linewidth]{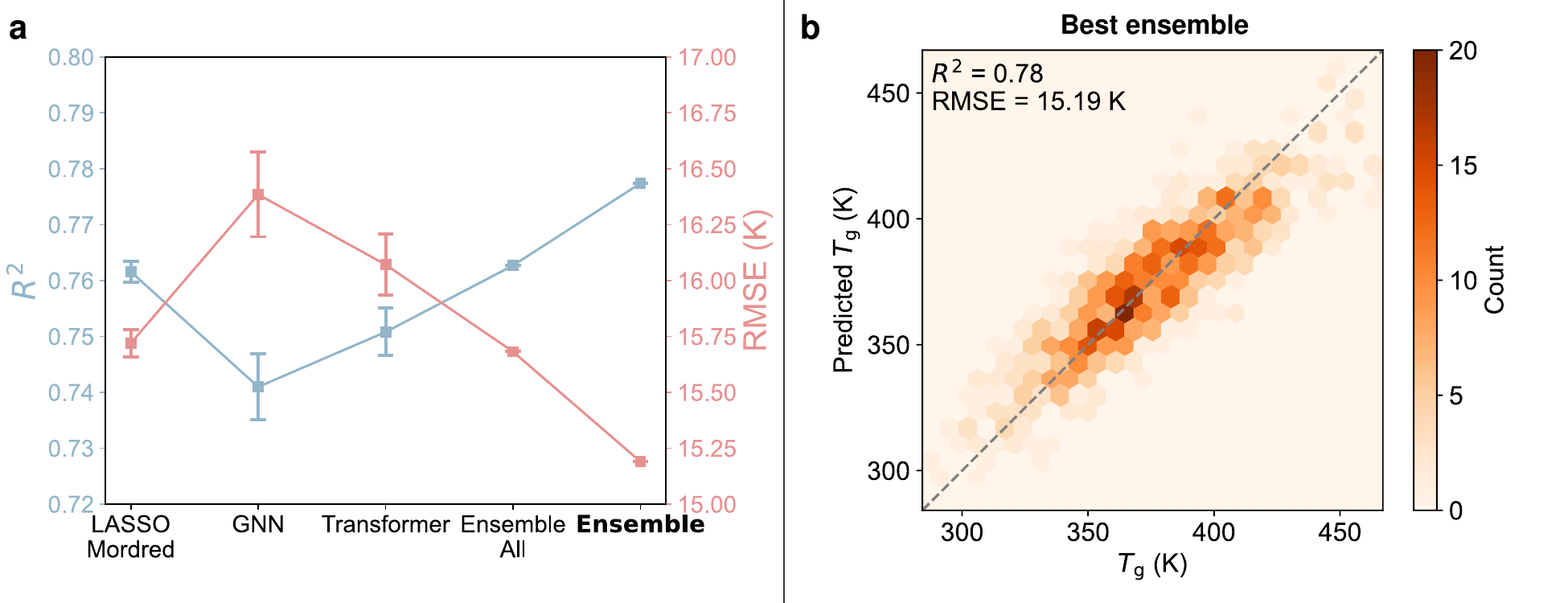}
\caption{Performance of other models and ensemble model in $T_\mathrm{g}$ prediction. (a) $R^2$ and RMSE between ground truth $T_\mathrm{g}$ and predicted $T_\mathrm{g}$ of LASSO\_Mordred, GNN, Transformer, ensemble of all models and the best ensemble model. (b) Predicted $T_\mathrm{g}$ vs. ground truth $T_\mathrm{g}$ of the best ensemble model. The best-performing model is highlighted in bold.}
\label{ensemble}
\end{figure}

The performance of LASSO\_Mordred is compared with GNN and Transformer models in Figure \ref{ensemble}a, where it also demonstrates superior prediction accuracy over the deep learning models. To further improve $T_\mathrm{g}$ prediction, we implement a model ensemble approach by averaging the predicted $T_\mathrm{g}$ values from multiple models. We find that an ensemble consisting of XGBoost with fingerprints (XGBoost\_fp), XGBoost with Mordred descriptors (XGBoost\_Mordred), GNN, and Transformer models achieves the highest correlation with ground truth values in the test set, yielding an $R^2$ value of 0.78 and a RMSE of 15.19 K (Figure \ref{ensemble}b). In contrast, an ensemble incorporating all 22 models performs worse due to redundancy in overlapping features. Notably, the four models in the best-performing ensemble encompass four major types of feature representations: molecular fingerprints, molecular descriptors, graphs, and SMILES strings. These results suggest that averaging predictions from models with diverse feature representations leads to higher accuracy than relying on any single model alone, demonstrating the effectiveness of the ensemble learning approach.

\subsection{Interpretability offered by machine learning models}

\begin{figure}[t]
\centering
\includegraphics[width=\linewidth]{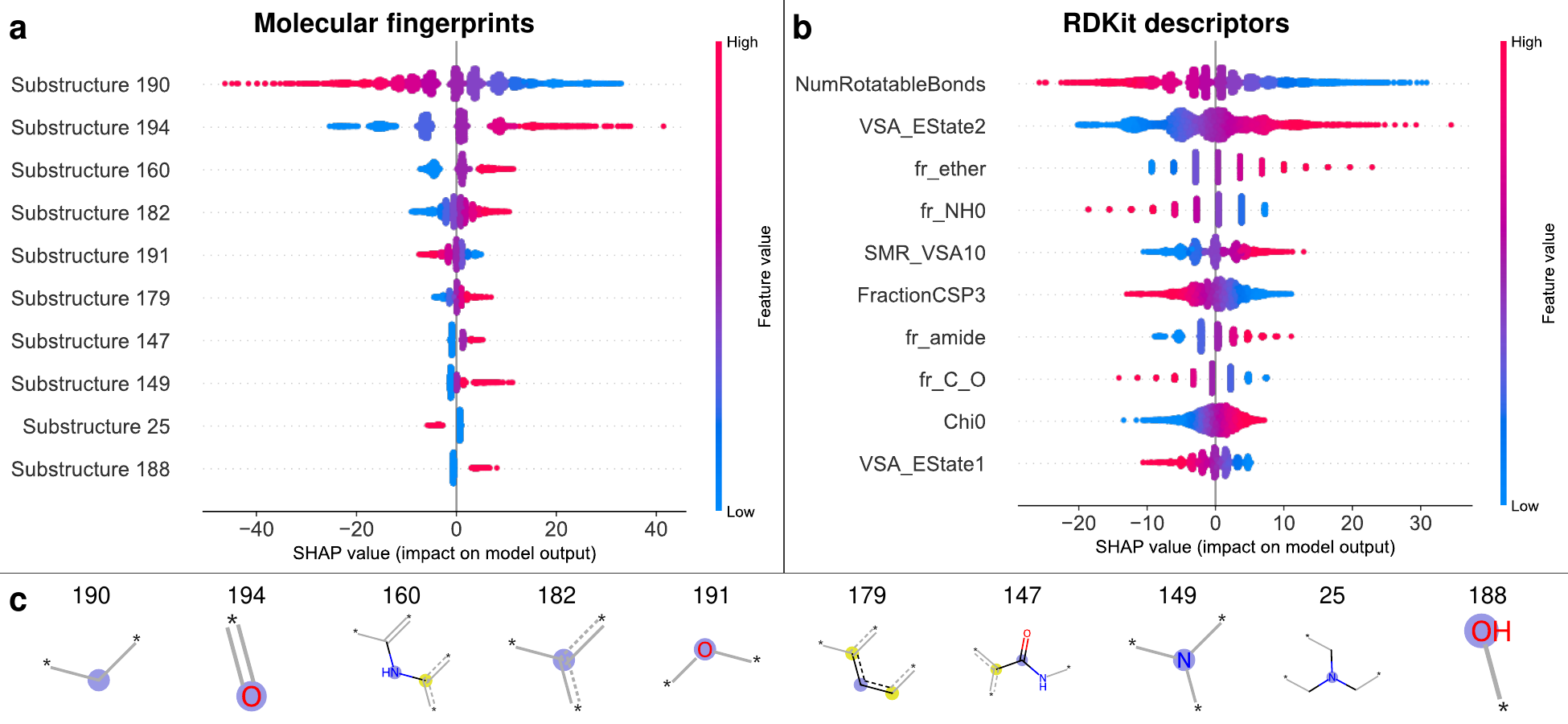}
\caption{Average SHAP values (feature importance scores) of ten most impactful (a) substructures and (b) RDKit descriptors calculated from the trained LASSO, RF, XGBoost and FFNN models. (c) Schematics of the substructures. The central atom of each substructure is highlighted in blue, while aromatic atoms are marked in yellow. The connectivity between atoms is represented in light gray.}
\label{shap}
\end{figure}

We perform SHAP analysis on the trained models (LASSO, RF, XGBoost, and FFNN) to gain physical insights into the structure–$T_\mathrm{g}$ relationships. For molecular fingerprints and molecular descriptors, we compute SHAP values for each feature (substructure or descriptor), which quantify the impact of individual features on the $T_\mathrm{g}$ of vitrimers. The ten most impactful substructures are presented in Figures \ref{shap}, ranked by their relative importance. Substructure 190, identified as the most impactful, negatively influences $T_\mathrm{g}$ due to its aliphatic nature which makes polymer chains more flexible. The positively impactful substructures can be classified into two groups. The first group, comprising substructures 160, 179, and 147, increases $T_\mathrm{g}$ due to their aromaticity and increased chain rigidity. The second group, including substructures 194, 160, 147, 149, and 188, contains oxygen and nitrogen atoms, which are involved in hydrogen bonding within the vitrimer network, leading to an increase in $T_\mathrm{g}$. These trends are consistent with previous experimental studies \cite{naito1993molecular, painter1991effect}. Furthermore, our findings align with the work of Tao et al. \cite{tao2021machine}, which identified similar substructures using a LASSO model trained on an experimental $T_\mathrm{g}$ dataset. This consistency with literature confirms the validity of our results.

The most impactful RDKit descriptors are shown in Figure \ref{shap}b, with their definitions provided in Table S1 (Supporting Information). The descriptor with the most negative impact, the number of rotatable bonds, decreases vitrimer chain rigidity, leading to a negative correlation with $T_\mathrm{g}$. The most positively correlated descriptor, VSA\_Estate2, is a hybrid descriptor that captures the combined effect of van der Waals surface area and the electrotopological states of atoms. Another notable descriptor, FractionCSP3, represents the fraction of carbon atoms that are sp\textsuperscript{3}-hybridized and negatively impacts $T_\mathrm{g}$ due to a more aliphatic and flexible molecular structure. Several identified descriptors have also been reported in a previous study based on an experimental dataset \cite{babbar2024explainability}, further validating the effectiveness of our analysis. The most important Mordred descriptors, along with their descriptions, are presented in Figure S3 and Table S2 (Supporting Information). Similar to the RDKit descriptors, the most significant Mordred descriptor is the number of rotatable bonds, demonstrating consistency across different models and providing cross-validation within this study.

The insights derived from molecular fingerprints and descriptors highlight two primary factors influencing vitrimer $T_\mathrm{g}$ across different scales: chain mobility and molecular interactions. At the polymer chain scale, increased chain flexibility, facilitated by rotatable bonds and aliphatic structures, enhances conformational freedom and segmental motion, resulting in lower $T_\mathrm{g}$. In contrast, the presence of aromatic rings and bulky side groups contributes to greater chain rigidity, restricting mobility and leading to higher $T_\mathrm{g}$. At the atomistic scale, molecular interactions, such as hydrogen bonding and van der Waals forces, strengthen local interactions within the vitrimer network, increasing energy barriers to segmental motion and thereby elevating $T_\mathrm{g}$. These findings offer valuable molecular design guidelines that can inform the discovery of vitrimers with tailored thermal properties.

\subsection{Discovering novel hypothetical vitrimers by virtual screening}

\begin{figure}[t]
\centering
\includegraphics[width=\linewidth]{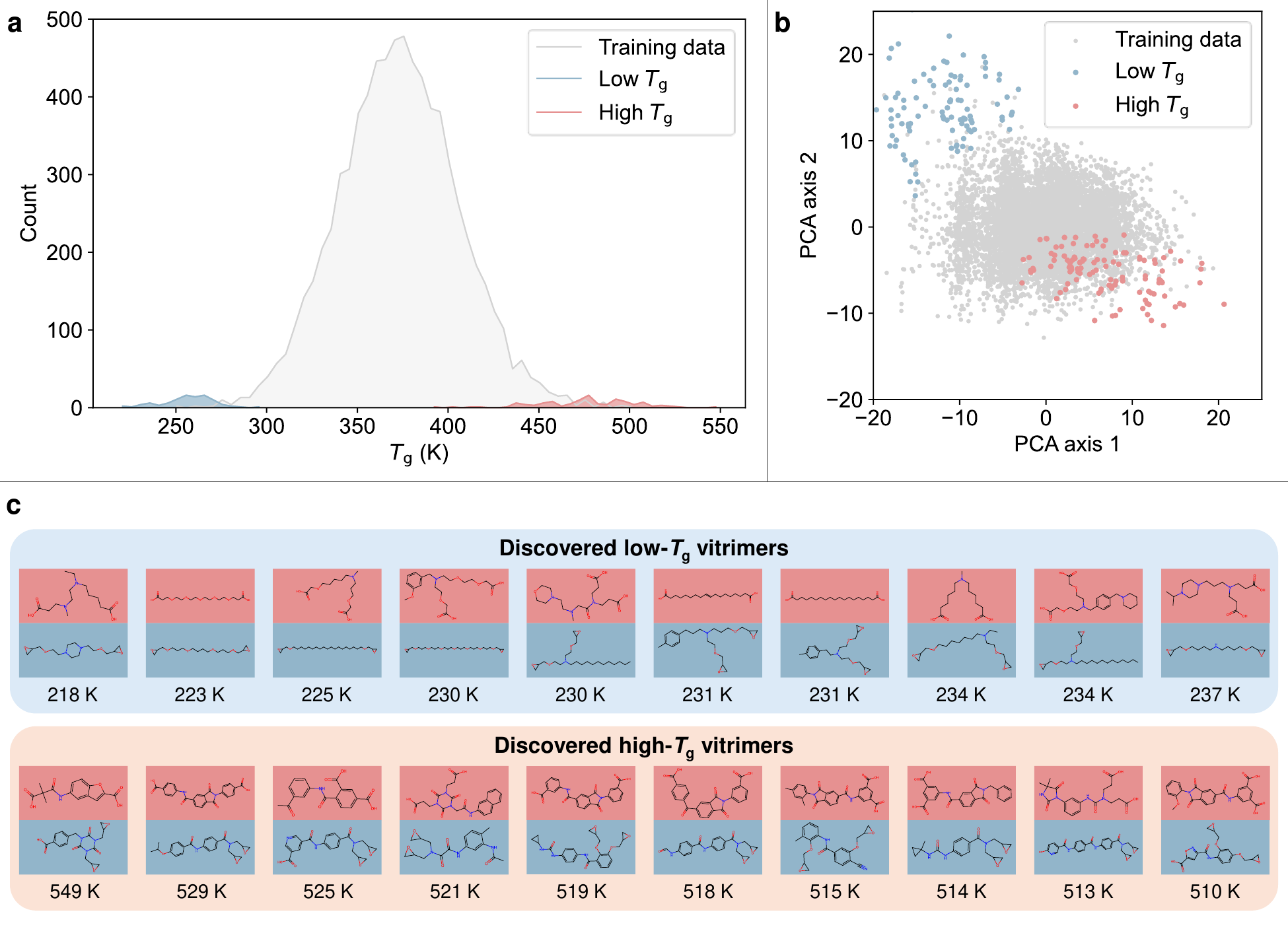}
\caption{Virtual screening of the hypothetical unlabeled dataset (see Table \ref{dataset}) and examples of discovered vitrimers whose $T_\mathrm{g}$ is validated by MD simulations. (a) $T_\mathrm{g}$ distribution of the training data and discovered vitrimers. (b) Chemical space of the training data and discovered vitrimers visualized by PCA of molecular fingerprints. (c) Molecular structures of discovered vitrimers with high and low $T_\mathrm{g}$. All presented $T_\mathrm{g}$ values are obtained from MD simulations.}
\label{screening}
\end{figure}

After identifying the ensemble model with the best performance on the test set, we proceed with predicting $T_\mathrm{g}$ for all vitrimers in the hypothetical unlabeled dataset (see Table \ref{dataset}). Approximately 100 vitrimers with the highest and lowest predicted $T_\mathrm{g}$ values are selected for validation through MD simulations and calibration. The simulation and calibration protocols remain consistent with those used in generating the labeled dataset (see Supporting Information for details). The discovered high-$T_\mathrm{g}$ and low-$T_\mathrm{g}$ vitrimers successfully expand the property space of the training set (Figure \ref{screening}a), potentially broadening the applicability of vitrimers for use in more extreme environments. To further analyze the distribution of the discovered vitrimers relative to the training data, we apply principal component analysis (PCA) to reduce the molecular fingerprint representations to two-dimensional vectors. As shown in Figure \ref{screening}b, the novel vitrimers extend the chemical space beyond the training regime. The 10 vitrimers with the highest and lowest $T_\mathrm{g}$ are shown in Figure \ref{screening} where all presented $T_\mathrm{g}$ values are obtained from MD simulations and GP calibration. The novel low-$T_\mathrm{g}$ vitrimers exhibit a range from 218 K to 237 K, while the high-$T_\mathrm{g}$ vitrimers range from 510 K to 549 K (Figure \ref{screening}c). These results demonstrate that ML-assisted virtual screening enables the identification of novel vitrimers with desirable properties at a significantly lower cost compared to conventional experimental and simulation-based approaches.

The most impactful substructure frequencies and descriptors (see Figure \ref{shap}) of the discovered vitrimers are presented in Figure S4 (Supporting Information). The occurrence of positively correlated substructures and descriptors is higher in high-$T_\mathrm{g}$ vitrimers, while negatively correlated features are more prevalent in low-$T_\mathrm{g}$ vitrimers. This consistency cross-validates our feature importance analysis, demonstrating that the key factors identified in the training data also apply to novel vitrimers outside the training dataset. These findings further support the universality and robustness of the interpretable insights offered by our models.

\subsection{Experimental synthesis and characterization of discovered vitrimers}

\begin{figure}[!ht]
\centering
\includegraphics[width=\linewidth]{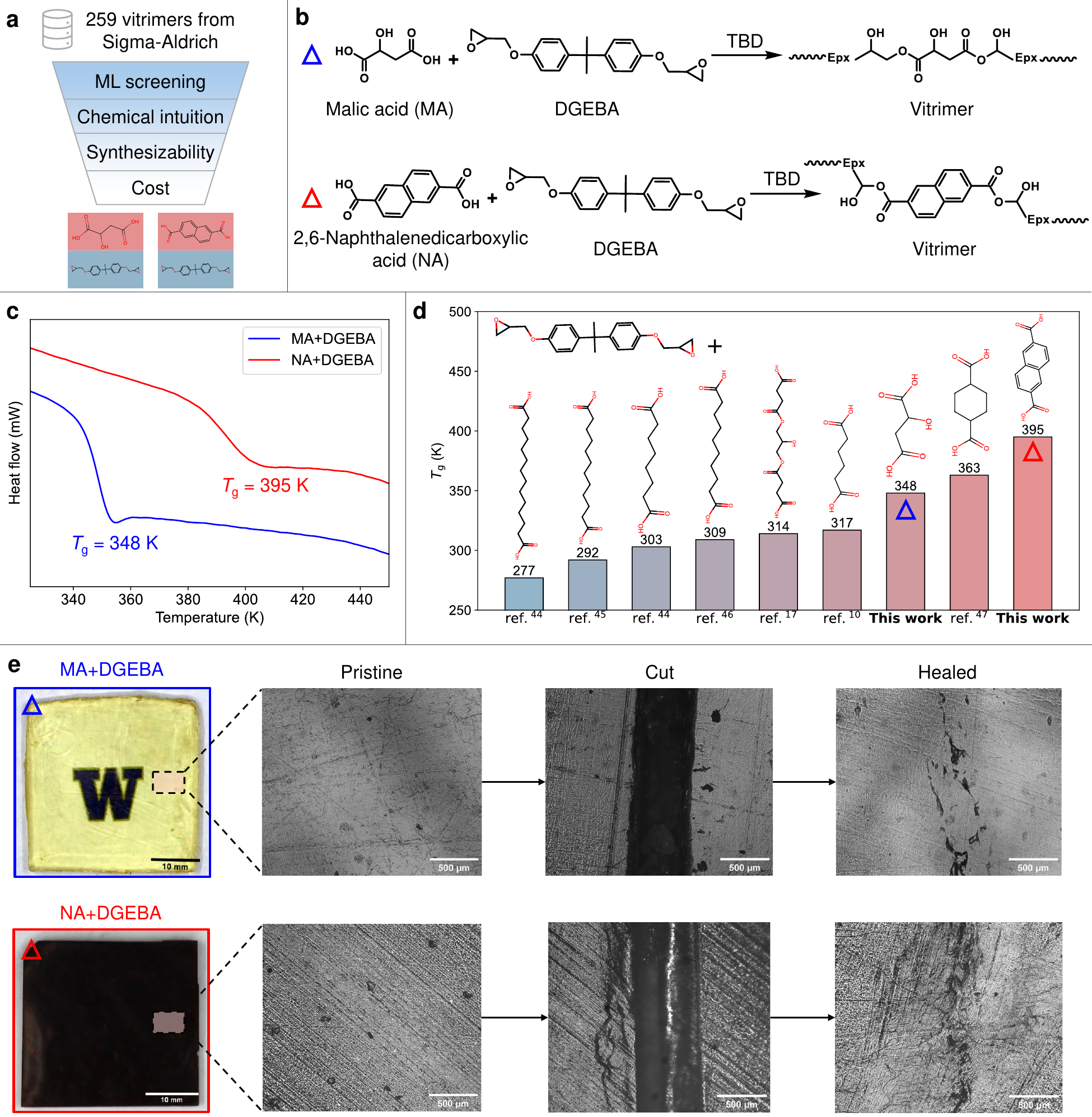}
\caption{Experimental synthesis and characterization of the discovered novel vitrimers. (a) The trained ensemble model is employed to virtually screen 259 vitrimers sourced from the Sigma-Aldrich database \cite{SigmaAldrich}. By integrating chemical intuition, we evaluate their synthesizability and cost, ultimately selecting two vitrimers for experimental synthesis and characterization. (b) Synthesis schemes of the vitrimers. (c) DSC results to measure the $T_\mathrm{g}$ of the synthesized vitrimers. (d) $T_\mathrm{g}$ of the vitrimers synthesized in this work and bifunctional transesterification vitrimers present in the literature \cite{shen2024influence, kaiser2020crucial, hubbard2021vitrimer, zheng2025ai, kamble2022reversing, tangthana2021epoxy}. (e) Images of pristine, cut and healed vitrimer specimens, demonstrating healability of the synthesized vitrimers in this work.}
\label{synthesis}
\end{figure}

To validate the predictive capability and practical relevance of our trained ensemble ML model, we transition from virtual screening to experimental synthesis. The ensemble model is used to efficiently evaluate the $T_\mathrm{g}$ of 259 vitrimer candidates composed of commercially available acids and epoxides from the Sigma-Aldrich database \cite{SigmaAldrich}. Guided by the model's $T_\mathrm{g}$ predictions, we identify promising candidates exhibiting high thermal performance. These predictions are further filtered based on chemical intuition, ease of synthesis and purchasing, and material cost, ultimately leading to the selection of two vitrimer systems for experimental validation (Figure \ref{synthesis}a).

Among the screened candidates, vitrimers based on D,L-malic acid (MA) and 2,6-naphthalenedicarboxylic acid (NA) paired with diglycidyl ether (DGEBA) are selected for synthesis, as they exhibit desired $T_\mathrm{g}$ values from the ML ensemble model. These acids are associated with structural features—such as aromaticity and limited rotatable bonds in NA and hydroxyl-rich backbone in MA—that are identified in our SHAP analysis as positively contributing to $T_\mathrm{g}$, reinforcing the rationale for their selection. The epoxide and acids are cured in the presence of the catalyst triazabicyclodecene (TBD) for vitrimer synthesis (Figure \ref{synthesis}b)

Thermal analysis through DSC measurements reveals $T_\mathrm{g}$ values of 348 K (75 \textdegree C) and 395 K (122 \textdegree C) for the MA-based and NA-based vitrimers, respectively (Figure \ref{synthesis}c). Notably, the NA-based vitrimer exhibits a higher $T_\mathrm{g}$ than any bifunctional transesterification vitrimer reported in the literature (see Table S3, Supporting Information) to the best of our knowledge \cite{shen2024influence, kaiser2020crucial, hubbard2021vitrimer, zheng2025ai, kamble2022reversing, tangthana2021epoxy}, as shown in Figure \ref{synthesis}d. The MA-based vitrimer also demonstrates a relatively higher $T_\mathrm{g}$ compared to most existing bifunctional transesterification vitrimers. The experimental $T_\mathrm{g}$ of both vitrimers agrees well with the ML-predicted $T_\mathrm{g}$ (364 and 390 K), indicating the accuracy of our ML model in predicting the actual $T_\mathrm{g}$ of vitrimers. To evaluate the healability of the synthesized vitrimers, pristine specimens are cut and subsequently healed at elevated temperatures. Microscopic examination of the pristine, cut, and healed samples (Figure \ref{synthesis}e) demonstrates complete damage recovery, confirming the healability of these novel vitrimer chemistries. The combination of high $T_\mathrm{g}$ and self-healing capability validates the effectiveness of our framework in identifying vitrimers with exceptional properties, thereby broadening the scope of potential vitrimer applications.

\section{Conclusions}

This work presents an integrated MD-ML virtual screening framework for discovering bifunctional transesterification vitrimers. Leveraging MD-generated data from our previous study \cite{zheng2025ai}, we systematically evaluate the predictive performance of ML models for vitrimer $T_\mathrm{g}$. Seven ML models are benchmarked across six molecular representations. Our findings demonstrate that an ensemble learning approach, which averages predictions from multiple models, outperforms individual models in $T_\mathrm{g}$ prediction. Furthermore, our feature importance analysis provides interpretability, identifying key structural and chemical factors that govern vitrimer $T_\mathrm{g}$. Using the trained ensemble model, we screen an unlabeled dataset of approximately 1 million hypothetical vitrimers and discover candidates with extreme $T_\mathrm{g}$ values beyond the training domain. We perform an additional screening of vitrimers composed of commercially available carboxylic acids and epoxides, identifying promising candidates for experimental evaluation. Two novel vitrimers are synthesized and characterized, confirming their exceptional properties. Both vitrimers exhibit relatively high $T_\mathrm{g}$, with one surpassing the $T_\mathrm{g}$ of any bifunctional transesterification vitrimer reported in the literature. This study highlights the feasibility of using MD-generated data to train ML models when experimental data is limited and successfully demonstrates the application of this approach in designing and validating novel vitrimers in experiments. Although this work focuses on bifunctional transesterification vitrimers, the presented approach can be readily adapted to other types of vitrimers and polymers with minimal modifications to the ML model architecture. The proposed framework provides an effective and efficient tool for polymer chemists to design synthesizable polymers with tailored properties, expanding the potential applications of vitrimers.

\section*{Author contributions}

\textbf{Yiwen Zheng}: Conceptualization, Data curation, Formal analysis, Investigation, Methodology, Software, Validation, Visualization, Writing – original draft, Writing – review and editing. \textbf{Agni K. Biswal}: Conceptualization, Data curation, Formal analysis, Investigation, Methodology, Visualization, Writing – original draft, Writing – review and editing. \textbf{Yaqi Guo}: Methodology, Software, Writing – review and editing. \textbf{Prakash Thakolkaran}: Methodology, Software, Writing – review and editing. \textbf{Yash Kokane}: Methodology, Software, Writing – review and editing. \textbf{Vikas Varshney}: Methodology, Supervision, Writing – review and editing. \textbf{Siddhant Kumar}: Methodology, Supervision, Writing – review and editing. \textbf{Aniruddh Vashisth}: Conceptualization, Funding acquisition, Methodology, Project administration, Resources, Supervision, Writing – review and editing.

\section*{Declaration of Competing Interest}

The authors declare no competing interests.

\section*{Data and code availability}

The data and code that support the findings of this study are openly available at \url{https://github.com/vashisth-lab/VitrimerScreening}.

\section*{Acknowledgments}

This work was completed using the advanced computational, storage, and networking infrastructure provided by the Hyak supercomputer system at the University of Washington, Seattle. YZ would like to thank Clean Energy Institute (CEI) Fellowship for financial support. This work was partially funded by Amazon Research Awards and National Science Foundation (NSF) (Award Number 2421235).

\bibliographystyle{elsarticle-num}
\bibliography{reference}

\begin{thebibliography}{10}
\expandafter\ifx\csname url\endcsname\relax
  \def\url#1{\texttt{#1}}\fi
\expandafter\ifx\csname urlprefix\endcsname\relax\def\urlprefix{URL }\fi
\expandafter\ifx\csname href\endcsname\relax
  \def\href#1#2{#2} \def\path#1{#1}\fi

\bibitem{young2011introduction}
R.~J. Young, P.~A. Lovell, Introduction to polymers, CRC press, 2011.

\bibitem{montarnal2011silica}
D.~Montarnal, M.~Capelot, F.~Tournilhac, L.~Leibler, Silica-like malleable materials from permanent organic networks, Science 334~(6058) (2011) 965--968.

\bibitem{capelot2012catalytic}
M.~Capelot, M.~M. Unterlass, F.~Tournilhac, L.~Leibler, Catalytic control of the vitrimer glass transition, ACS Macro Letters 1~(7) (2012) 789--792.

\bibitem{jin2019malleable}
Y.~Jin, Z.~Lei, P.~Taynton, S.~Huang, W.~Zhang, Malleable and recyclable thermosets: the next generation of plastics, Matter 1~(6) (2019) 1456--1493.

\bibitem{krishnakumar2020vitrimers}
B.~Krishnakumar, R.~P. Sanka, W.~H. Binder, V.~Parthasarthy, S.~Rana, N.~Karak, Vitrimers: Associative dynamic covalent adaptive networks in thermoset polymers, Chemical Engineering Journal 385 (2020) 123820.

\bibitem{hoogeboom2022equivariant}
E.~Hoogeboom, V.~G. Satorras, C.~Vignac, M.~Welling, Equivariant diffusion for molecule generation in 3d, in: International conference on machine learning, PMLR, 2022, pp. 8867--8887.

\bibitem{hansson2002molecular}
T.~Hansson, C.~Oostenbrink, W.~van Gunsteren, Molecular dynamics simulations, Current opinion in structural biology 12~(2) (2002) 190--196.

\bibitem{vashisth2018accelerated}
A.~Vashisth, C.~Ashraf, W.~Zhang, C.~E. Bakis, A.~C. Van~Duin, Accelerated reaxff simulations for describing the reactive cross-linking of polymers, The Journal of Physical Chemistry A 122~(32) (2018) 6633--6642.

\bibitem{yu2001polymer}
K.-q. Yu, Z.-s. Li, J.~Sun, Polymer structures and glass transition: A molecular dynamics simulation study, Macromolecular theory and simulations 10~(6) (2001) 624--633.

\bibitem{kamble2022reversing}
M.~Kamble, A.~Vashisth, H.~Yang, S.~Pranompont, C.~R. Picu, D.~Wang, N.~Koratkar, Reversing fatigue in carbon-fiber reinforced vitrimer composites, Carbon 187 (2022) 108--114.

\bibitem{kranenburg2009challenges}
J.~M. Kranenburg, C.~A. Tweedie, K.~J. van Vliet, U.~S. Schubert, Challenges and progress in high-throughput screening of polymer mechanical properties by indentation, Advanced Materials 21~(35) (2009) 3551--3561.

\bibitem{tao2021machine}
L.~Tao, G.~Chen, Y.~Li, Machine learning discovery of high-temperature polymers, Patterns 2~(4) (2021).

\bibitem{thakolkaran2024deep}
P.~Thakolkaran, Y.~Zheng, Y.~Guo, A.~Vashisth, S.~Kumar, Deep learning reveals key predictors of thermal conductivity in covalent organic frameworks, arXiv preprint arXiv:2409.06457 (2024).

\bibitem{batra2020polymers}
R.~Batra, H.~Dai, T.~D. Huan, L.~Chen, C.~Kim, W.~R. Gutekunst, L.~Song, R.~Ramprasad, Polymers for extreme conditions designed using syntax-directed variational autoencoders, Chemistry of Materials 32~(24) (2020) 10489--10500.

\bibitem{yang2022machine}
J.~Yang, L.~Tao, J.~He, J.~R. McCutcheon, Y.~Li, Machine learning enables interpretable discovery of innovative polymers for gas separation membranes, Science Advances 8~(29) (2022) eabn9545.

\bibitem{otsuka2011polyinfo}
S.~Otsuka, I.~Kuwajima, J.~Hosoya, Y.~Xu, M.~Yamazaki, Polyinfo: Polymer database for polymeric materials design, in: 2011 International Conference on Emerging Intelligent Data and Web Technologies, IEEE, 2011, pp. 22--29.

\bibitem{zheng2025ai}
Y.~Zheng, P.~Thakolkaran, A.~K. Biswal, J.~A. Smith, Z.~Lu, S.~Zheng, B.~H. Nguyen, S.~Kumar, A.~Vashisth, Ai-guided inverse design and discovery of recyclable vitrimeric polymers, Advanced Science 12~(6) (2025) 2411385.

\bibitem{tao2023machine}
L.~Tao, J.~He, T.~Arbaugh, J.~R. McCutcheon, Y.~Li, Machine learning prediction on the fractional free volume of polymer membranes, Journal of Membrane Science 665 (2023) 121131.

\bibitem{xie2022cloud}
T.~Xie, H.-K. Kwon, D.~Schweigert, S.~Gong, A.~France-Lanord, A.~Khajeh, E.~Crabb, M.~Puzon, C.~Fajardo, W.~Powelson, et~al., A cloud platform for automating and sharing analysis of raw simulation data from high throughput polymer molecular dynamics simulations, arXiv preprint arXiv:2208.01692 (2022).

\bibitem{rdkit}
G.~Landrum, P.~Tosco, B.~Kelley, Ric, D.~Cosgrove, sriniker, gedeck, R.~Vianello, NadineSchneider, E.~Kawashima, D.~N, G.~Jones, A.~Dalke, B.~Cole, M.~Swain, S.~Turk, AlexanderSavelyev, A.~Vaucher, M.~Wójcikowski, I.~Take, D.~Probst, K.~Ujihara, V.~F. Scalfani, guillaume godin, J.~Lehtivarjo, A.~Pahl, R.~Walker, F.~Berenger, jasondbiggs, strets123, Rdkit: Open-source cheminformatics., \url{https://www.rdkit.org}.

\bibitem{moriwaki2018mordred}
H.~Moriwaki, Y.-S. Tian, N.~Kawashita, T.~Takagi, Mordred: a molecular descriptor calculator, Journal of cheminformatics 10~(1) (2018) 1--14.

\bibitem{rogers2010extended}
D.~Rogers, M.~Hahn, Extended-connectivity fingerprints, Journal of chemical information and modeling 50~(5) (2010) 742--754.

\bibitem{weininger1988smiles}
D.~Weininger, Smiles, a chemical language and information system. 1. introduction to methodology and encoding rules, Journal of chemical information and computer sciences 28~(1) (1988) 31--36.

\bibitem{wu2020comprehensive}
Z.~Wu, S.~Pan, F.~Chen, G.~Long, C.~Zhang, S.~Y. Philip, A comprehensive survey on graph neural networks, IEEE transactions on neural networks and learning systems 32~(1) (2020) 4--24.

\bibitem{vaswani2017attention}
A.~Vaswani, N.~Shazeer, N.~Parmar, J.~Uszkoreit, L.~Jones, A.~N. Gomez, {\L}.~Kaiser, I.~Polosukhin, Attention is all you need, Advances in neural information processing systems 30 (2017).

\bibitem{yan2023overcoming}
C.~Yan, X.~Feng, J.~Konlan, P.~Mensah, G.~Li, Overcoming the barrier: designing novel thermally robust shape memory vitrimers by establishing a new machine learning framework, Physical Chemistry Chemical Physics 25~(43) (2023) 30049--30065.

\bibitem{jaeger2018mol2vec}
S.~Jaeger, S.~Fulle, S.~Turk, Mol2vec: unsupervised machine learning approach with chemical intuition, Journal of chemical information and modeling 58~(1) (2018) 27--35.

\bibitem{sterling2015zinc}
T.~Sterling, J.~J. Irwin, Zinc 15--ligand discovery for everyone, Journal of chemical information and modeling 55~(11) (2015) 2324--2337.

\bibitem{SigmaAldrich}
{Sigma-Aldrich}, \url{https://www.sigmaaldrich.com/US/en}, accessed: February 2025.

\bibitem{morgan1965generation}
H.~L. Morgan, The generation of a unique machine description for chemical structures-a technique developed at chemical abstracts service., Journal of chemical documentation 5~(2) (1965) 107--113.

\bibitem{tao2021benchmarking}
L.~Tao, V.~Varshney, Y.~Li, Benchmarking machine learning models for polymer informatics: an example of glass transition temperature, Journal of Chemical Information and Modeling 61~(11) (2021) 5395--5413.

\bibitem{hickey2024applying}
K.~Hickey, J.~Feinstein, G.~Sivaraman, M.~MacDonell, E.~Yan, C.~Matherson, S.~Coia, J.~Xu, K.~Picel, Applying machine learning and quantum chemistry to predict the glass transition temperatures of polymers, Computational Materials Science 238 (2024) 112933.

\bibitem{pedregosa2011scikit}
F.~Pedregosa, G.~Varoquaux, A.~Gramfort, V.~Michel, B.~Thirion, O.~Grisel, M.~Blondel, P.~Prettenhofer, R.~Weiss, V.~Dubourg, et~al., Scikit-learn: Machine learning in python, the Journal of machine Learning research 12 (2011) 2825--2830.

\bibitem{chen2016xgboost}
T.~Chen, C.~Guestrin, Xgboost: A scalable tree boosting system, in: Proceedings of the 22nd acm sigkdd international conference on knowledge discovery and data mining, 2016, pp. 785--794.

\bibitem{paszke2019pytorch}
A.~Paszke, S.~Gross, F.~Massa, A.~Lerer, J.~Bradbury, G.~Chanan, T.~Killeen, Z.~Lin, N.~Gimelshein, L.~Antiga, et~al., Pytorch: An imperative style, high-performance deep learning library, Advances in neural information processing systems 32 (2019).

\bibitem{xu2023transpolymer}
C.~Xu, Y.~Wang, A.~Barati~Farimani, Transpolymer: a transformer-based language model for polymer property predictions, npj Computational Materials 9~(1) (2023) 64.

\bibitem{alfaraj2023model}
Y.~S. AlFaraj, S.~Mohapatra, P.~Shieh, K.~E. Husted, D.~G. Ivanoff, E.~M. Lloyd, J.~C. Cooper, Y.~Dai, A.~P. Singhal, J.~S. Moore, et~al., A model ensemble approach enables data-driven property prediction for chemically deconstructable thermosets in the low-data regime, ACS Central Science 9~(9) (2023) 1810--1819.

\bibitem{esmaeili2023accelerated}
H.~Esmaeili, R.~Rizvi, An accelerated strategy to characterize mechanical properties of polymer composites using the ensemble learning approach, Computational Materials Science 229 (2023) 112432.

\bibitem{gao2024accurately}
W.~Gao, H.~Wang, Y.~Xu, Y.~Yang, Q.~Gu, X.~Duan, B.~Wang, X.~Zhou, Accurately predicting multiple performance of 3d printing photopolymers using ensemble learning, ACS Applied Polymer Materials 6~(8) (2024) 4501--4508.

\bibitem{lundberg2017unified}
S.~Lundberg, A unified approach to interpreting model predictions, arXiv preprint arXiv:1705.07874 (2017).

\bibitem{naito1993molecular}
K.~Naito, A.~Miura, Molecular design for nonpolymeric organic dye glasses with thermal stability: relations between thermodynamic parameters and amorphous properties, The Journal of Physical Chemistry 97~(23) (1993) 6240--6248.

\bibitem{painter1991effect}
P.~C. Painter, J.~F. Graf, M.~M. Coleman, Effect of hydrogen bonding on the enthalpy of mixing and the composition dependence of the glass transition temperature in polymer blends, Macromolecules 24~(20) (1991) 5630--5638.

\bibitem{babbar2024explainability}
A.~Babbar, S.~Ragunathan, D.~Mitra, A.~Dutta, T.~K. Patra, Explainability and extrapolation of machine learning models for predicting the glass transition temperature of polymers, Journal of Polymer Science 62~(6) (2024) 1175--1186.

\bibitem{shen2024influence}
S.~Shen, V.~K. Thakur, A.~A. Skordos, Influence of monomer structure and catalyst concentration on topological transition and dynamic properties of dicarboxylic acid-epoxy vitrimers, Journal of Applied Polymer Science 141~(40) (2024) e56028.

\bibitem{kaiser2020crucial}
S.~Kaiser, P.~Novak, M.~Giebler, M.~Gschwandl, P.~Novak, G.~Pilz, M.~Morak, S.~Schl{\"o}gl, The crucial role of external force in the estimation of the topology freezing transition temperature of vitrimers by elongational creep measurements, Polymer 204 (2020) 122804.

\bibitem{hubbard2021vitrimer}
A.~M. Hubbard, Y.~Ren, D.~Konkolewicz, A.~Sarvestani, C.~R. Picu, G.~S. Kedziora, A.~Roy, V.~Varshney, D.~Nepal, Vitrimer transition temperature identification: coupling various thermomechanical methodologies, ACS Applied Polymer Materials 3~(4) (2021) 1756--1766.

\bibitem{tangthana2021epoxy}
K.~Tangthana-Umrung, Q.~A. Poutrel, M.~Gresil, Epoxy homopolymerization as a tool to tune the thermo-mechanical properties and fracture toughness of vitrimers, Macromolecules 54~(18) (2021) 8393--8406.

\end{thebibliography}


\begin{thebibliography}{10}
\expandafter\ifx\csname url\endcsname\relax
  \def\url#1{\texttt{#1}}\fi
\expandafter\ifx\csname urlprefix\endcsname\relax\def\urlprefix{URL }\fi
\expandafter\ifx\csname href\endcsname\relax
  \def\href#1#2{#2} \def\path#1{#1}\fi

\bibitem{zheng2025ai}
Y.~Zheng, P.~Thakolkaran, A.~K. Biswal, J.~A. Smith, Z.~Lu, S.~Zheng, B.~H. Nguyen, S.~Kumar, A.~Vashisth, Ai-guided inverse design and discovery of recyclable vitrimeric polymers, Advanced Science 12~(6) (2025) 2411385.

\bibitem{sterling2015zinc}
T.~Sterling, J.~J. Irwin, Zinc 15--ligand discovery for everyone, Journal of chemical information and modeling 55~(11) (2015) 2324--2337.

\bibitem{SigmaAldrich}
{Sigma-Aldrich}, \url{https://www.sigmaaldrich.com/US/en}, accessed: February 2025.

\bibitem{hickey2024applying}
K.~Hickey, J.~Feinstein, G.~Sivaraman, M.~MacDonell, E.~Yan, C.~Matherson, S.~Coia, J.~Xu, K.~Picel, Applying machine learning and quantum chemistry to predict the glass transition temperatures of polymers, Computational Materials Science 238 (2024) 112933.

\bibitem{xu2023transpolymer}
C.~Xu, Y.~Wang, A.~Barati~Farimani, Transpolymer: a transformer-based language model for polymer property predictions, npj Computational Materials 9~(1) (2023) 64.

\bibitem{shen2024influence}
S.~Shen, V.~K. Thakur, A.~A. Skordos, Influence of monomer structure and catalyst concentration on topological transition and dynamic properties of dicarboxylic acid-epoxy vitrimers, Journal of Applied Polymer Science 141~(40) (2024) e56028.

\bibitem{kaiser2020crucial}
S.~Kaiser, P.~Novak, M.~Giebler, M.~Gschwandl, P.~Novak, G.~Pilz, M.~Morak, S.~Schl{\"o}gl, The crucial role of external force in the estimation of the topology freezing transition temperature of vitrimers by elongational creep measurements, Polymer 204 (2020) 122804.

\bibitem{hubbard2021vitrimer}
A.~M. Hubbard, Y.~Ren, D.~Konkolewicz, A.~Sarvestani, C.~R. Picu, G.~S. Kedziora, A.~Roy, V.~Varshney, D.~Nepal, Vitrimer transition temperature identification: coupling various thermomechanical methodologies, ACS Applied Polymer Materials 3~(4) (2021) 1756--1766.

\bibitem{kamble2022reversing}
M.~Kamble, A.~Vashisth, H.~Yang, S.~Pranompont, C.~R. Picu, D.~Wang, N.~Koratkar, Reversing fatigue in carbon-fiber reinforced vitrimer composites, Carbon 187 (2022) 108--114.

\bibitem{tangthana2021epoxy}
K.~Tangthana-Umrung, Q.~A. Poutrel, M.~Gresil, Epoxy homopolymerization as a tool to tune the thermo-mechanical properties and fracture toughness of vitrimers, Macromolecules 54~(18) (2021) 8393--8406.

\end{thebibliography}

\end{document}


\begin{frontmatter}

\title{Toward Sustainable Polymer Design: A Molecular Dynamics-Informed Machine Learning Approach for Vitrimers \\ [0.3cm] \Large\textit{Supporting Information}}

\author[UW]{Yiwen Zheng}
\author[UW]{Agni K. Biswal}
\author[TUD]{Yaqi Guo}
\author[TUD]{Prakash Thakolkaran}
\author[UW]{Yash Kokane}
\author[AFRL]{Vikas Varshney}
\author[TUD]{Siddhant Kumar}
\author[UW]{Aniruddh Vashisth\corref{cor1}}

\cortext[cor1]{Email: vashisth@uw.edu}

\affiliation[UW]{organization={Department of Mechanical Engineering, University of Washington}, 
            city={Seattle},
            state={WA},
            country={USA}}
\affiliation[TUD]{organization={Department of Materials Science and Engineering, Delft University of Technology}, 
            city={Delft},
            country={The Netherlands}}
\affiliation[AFRL]{organization={Materials and Manufacturing Directorate, Air Force Research Laboratory}, 
            city={Wright-Patterson Air Force Base},
            state={OH},
            country={USA}}

\end{frontmatter}

\section{Vitrimer datasets and virtual screening}

In this work, we utilize the vitrimer dataset developed in our previous study \cite{zheng2025ai}, which consists of 1 million hypothetical vitrimers sampled from random combinations of 50,000 bifunctional carboxylic acid and 50,000 bifunctional epoxide molecules derived from the ZINC15 database \cite{sterling2015zinc}. To compute $T_\mathrm{g}$, MD simulations are performed on 8,424 vitrimers randomly sampled from the 1 million dataset. For each vitrimer composition, the acid and epoxide monomers are alternately connected to construct a polymer chain of approximately 1,000 atoms. The initial configuration consists of four such chains, which undergo energy minimization and annealing to eliminate local heterogeneities. The annealed system is then gradually cooled from 800 K to 100 K, with system density recorded at each temperature. The $T_\mathrm{g}$ value is determined as the intersection point of the bilinear fit of the density-temperature profile. To reduce uncertainty arising from the stochastic nature of MD simulations, five replicate simulations are conducted for each vitrimer, and the final $T_\mathrm{g}$ is obtained by averaging across these replicates.

Due to differences in time and length scales between MD simulations and real experiments, MD-calculated $T_\mathrm{g}$ values are typically overestimated compared to experimental measurements. To account for this artifact, the MD-calculated $T_\mathrm{g}$ is further calibrated against experimental values using a Gaussian Process (GP) model. The GP model is trained to predict the offset between experimental and MD-calculated $T_\mathrm{g}$ based on a polymer dataset with experimentally measured $T_\mathrm{g}$ values from the literature. Once trained, the model is applied to estimate the experimental-MD $T_\mathrm{g}$ offset for all 8,424 vitrimers, and the MD-calculated $T_\mathrm{g}$ values are subsequently calibrated to approximate experimental $T_\mathrm{g}$. The calibrated $T_\mathrm{g}$ serves as a proxy for experimental $T_\mathrm{g}$ and is used as the ground truth for ML model training in this work. More details on the MD simulations and GP calibration process can be found in our previous study \cite{zheng2025ai}.

The 8,424 vitrimers with their associated calibrated $T_\mathrm{g}$ values constitute the labeled dataset (Table 1 in the manuscript), which is used to train the ML models. The remaining 991,576 vitrimers form the hypothetical unlabeled dataset (Table 1 in the manuscript). Additionally, we construct a synthesizable unlabeled dataset (Table 1 in the manuscript) by selecting 37 commercially available bifunctional carboxylic acids and 7 bifunctional epoxides from the Sigma-Aldrich database \cite{SigmaAldrich}. Once the ML models are trained and the best-performing model is identified, we use it to predict the $T_\mathrm{g}$ values of vitrimers in both unlabeled datasets, facilitating the discovery of novel vitrimers with desirable properties. For vitrimers identified from the hypothetical unlabeled dataset, their $T_\mathrm{g}$ values are further validated using the previously described MD simulations and Gaussian Process (GP) calibration. For vitrimers selected from the synthesizable unlabeled dataset, we examine their synthesizability and select two candidates for experimental synthesis and characterization.

\section{Details on machine learning models}

In this work, we train and benchmark seven different ML models covering six feature representations, as shown in Figure \ref{models}. A brief introduction of each ML model is given below.

\begin{itemize}
    \item Least absolute shrinkage and selection operator (LASSO): LASSO is a regression technique that performs variable selection and regularization by imposing an L1 penalty on the regression coefficients, shrinking some to zero to improve model interpretability and prevent overfitting.
    \item Random forest (RF): RF is a machine learning method that constructs multiple decision trees during training and combines their outputs to improve prediction accuracy and reduce overfitting.
    \item Support vector regression (SVR): SVR is a machine learning technique based on support vector machines that fits a hyperplane or curve within a specified margin of tolerance to predict continuous outcomes, while minimizing model complexity and ensuring robustness to outliers.
    \item Extreme gradient boosting (XGBoost): XGBoost is a machine learning method based on gradient boosting, which builds multiple decision trees sequentially, optimizing residual errors and using techniques like regularization and parallel processing to enhance performance and prevent overfitting.
    \item Feedforward neural network (FFNN): FFNN is a type of artificial neural network where information flows from input to output layers through one or more hidden layers, without any cycles or loops.
    \item Graph neural network (GNN): GNN a type of neural network designed to process data represented as graphs by leveraging the graph structure, updating node embeddings through iterative message passing between nodes and edges.
    \item Transformer: A Transformer is a neural network architecture based on self-attention mechanisms, designed to process sequential data by capturing long-range dependencies and contextual relationships.
\end{itemize}

The FFNN models consist of hidden layers of equal dimensions with the ReLU activation function. The GNN model is implemented following the architecture by Hickey et al. \cite{hickey2024applying}, which involves alternating graph convolutional layers and fully connected hidden layers. The Transformer model is adapted from the pretrained TransPolymer model \cite{xu2023transpolymer} and fine-tuned using the vitrimer dataset. All models are trained and optimized using a five-fold cross-validation approach. For all models except the Transformer, hyperparameters are selected through 100 iterations of randomized search across the parameter space, aiming to minimize the average root mean square error (RMSE) on the validation set across all five folds. The hyperparameters of the Transformer model are determined by evaluating 12 combinations of the recommended hyperparameters in the original study \cite{xu2023transpolymer}. For deep learning models (FFNN, GNN, and Transformer), training is conducted for 100 epochs per fold, except for the Transformer, which is trained for 30 epochs. The best-performing model is selected based on the epoch achieving the lowest RMSE on the validation set. The trained models are subsequently used to predict $T_\mathrm{g}$ for the test set, and their performance is assessed using the coefficient of determination ($R^2$) and RMSE. Each performance metric is averaged over the five models obtained from the five cross-validation folds, and both the mean and standard deviation are reported, as shown in Figure 2 and Figure 3 of the manuscript.

Apart from individual models, we find that averaging predictions from multiple models improves prediction performance on the test set, a strategy commonly referred to as the ensemble learning approach. The predicted $T_\mathrm{g}$ values for the test set are averaged across 20 trained models, comprising five cross-validation folds each of XGBoost\_fp, XGBoost\_Mordred, GNN, and Transformer models, and compared against the ground truth $T_\mathrm{g}$ (Figure 3 in the manuscript). This ensemble model is further employed to predict $T_\mathrm{g}$ for the unlabeled datasets, facilitating the discovery of novel vitrimers with desirable properties.

\clearpage

\begin{table}[!ht]
    \centering
    \begin{tabular}{m{10em}|m{30em}}
        \Xhline{2\arrayrulewidth}
        \textbf{Descriptor} & \textbf{Definition} \\
        \hline
        NumRotatableBonds & Number of rotatable bonds \\
        \hline
        VSA\_EState2 & The contribution of electrotopological states of all atoms whose contribution to Van der Waals surface area lies between 4.78 and 5 \\
        \hline
        fr\_ether & Number of ether oxygens (including phenoxy) \\
        \hline
        fr\_NH0 & Number of tertiary amines \\
        \hline
        SMR\_VSA10 & The contribution of Van der Waals surface area of all atoms whose contribution to molecular refractivity is larger than 4 \\
        \hline
        FractionCSP3 & Fraction of carbon atoms that are sp\textsuperscript{3} hybridized \\
        \hline
        fr\_amide & Number of amides \\
        \hline
        fr\_C\_O & Number of carbonyl oxygen atoms \\
        \hline
        Chi0 & Sum of $1/\sqrt{d_i}$ over all heavy atoms $i$ with $d_i > 0$; $d_i$ is the number of heavy neighbor atoms \\
        \hline
        VSA\_EState1 & The contribution of electrotopological states of all atoms whose contribution to Van der Waals surface area is smaller than 4.78 \\
        \Xhline{2\arrayrulewidth}
    \end{tabular}
    \caption{Description of the RDKit descriptors presented in Figure 4 of the manuscript.}
    \label{def_rdkit}
\end{table}

\begin{table}[!ht]
    \centering
    \begin{tabular}{m{10em}|m{30em}}
        \Xhline{2\arrayrulewidth}
        \textbf{Descriptor} & \textbf{Definition} \\
        \hline
        RotRatio & Rotatable bonds ratio \\
        \hline
        nHBAcc & Number of hydrogen bond acceptors \\
        \hline
        SdO & Sum of electrotopological state indices of atom type dO \\
        \hline
        SLogP\_VSA10 & The contribution of Van der Waals surface area of all atoms whose contribution to LogP lies between 0.4 and 0.5 \\
        \hline
        AETA\_beta\_s & Averaged sigma contribution to valence electron mobile count \\
        \hline
        piPC6 & 6-ordered pi-path count (log scale) \\
        \hline
        MIC3 & 3-ordered modified information content \\
        \hline
        GATS2dv & Geary coefficient of lag 2 weighted by valence electrons \\
        \hline
        FilterItLogS & Solubility parameter \\
        \hline
        SpDiam\_Dt & Spectral diamiter from detourn matrix \\
        \Xhline{2\arrayrulewidth}
    \end{tabular}
    \caption{Description of the Mordred descriptors presented in Figure \ref{shap_mordred}.}
    \label{def_mordred}
\end{table}

\begin{table}[!ht]
    \centering
    \begin{tabular}{l|l|l|l}
        \Xhline{2\arrayrulewidth}
        \textbf{Acid} & \textbf{Epoxide} & $T_\mathrm{g}$ (K) & \textbf{Reference} \\
        \hline
        Tetradecanedioic acid & DGEBA & 277 & Shen et al. (2024) \cite{shen2024influence} \\
        \hline
        Dodecanedioic acid & DGEBA & 292 & Kaiser et al. (2020) \cite{kaiser2020crucial} \\
        \hline
        Suberic acid & DGEBA & 303 & Shen et al. (2024) \cite{shen2024influence} \\
        \hline
        Sebacic acid & DGEBA & 309 & Hubbard et al. (2021) \cite{hubbard2021vitrimer} \\
        \hline
        CHOA & DGEBA & 314 & Zheng et al. (2025) \cite{zheng2025ai} \\
        \hline
        Adipic acid & DGEBA & 317 & Kamble et al. (2022) \cite{kamble2022reversing} \\
        \hline
        CHDA & DGEBA & 363 & Tangthana-Umrung et al. (2021) \cite{tangthana2021epoxy} \\
        \Xhline{2\arrayrulewidth}
    \end{tabular}
    \caption{Glass transition temperature ($T_\mathrm{g}$) of bifunctional transesterification vitrimers in the literature. Abbreviations: DGEBA $\rightarrow$ bisphenol A diglycidyl ether; CHDA $\rightarrow$ 1,4-cyclohexane dicarboxylic acid; CHOA $\rightarrow$ 4-[3-(3-carboxypropanoyloxy)-2-hydroxypropoxy]-4-oxobutanoic acid.}
    \label{literature}
\end{table}

\clearpage

\begin{figure}[!ht]
\centering
\includegraphics[width=0.7\linewidth]{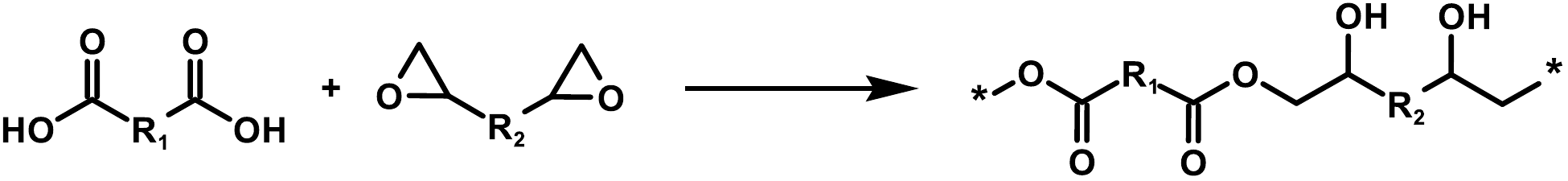}
\caption{Formation of vitrimer repeating unit by opening epoxide rings and reacting with carboxylic acids. The asterisk (*) symbol denotes polymer connection sites.}
\label{repeatunit}
\end{figure}

\begin{figure}[!ht]
\centering
\includegraphics[width=\linewidth]{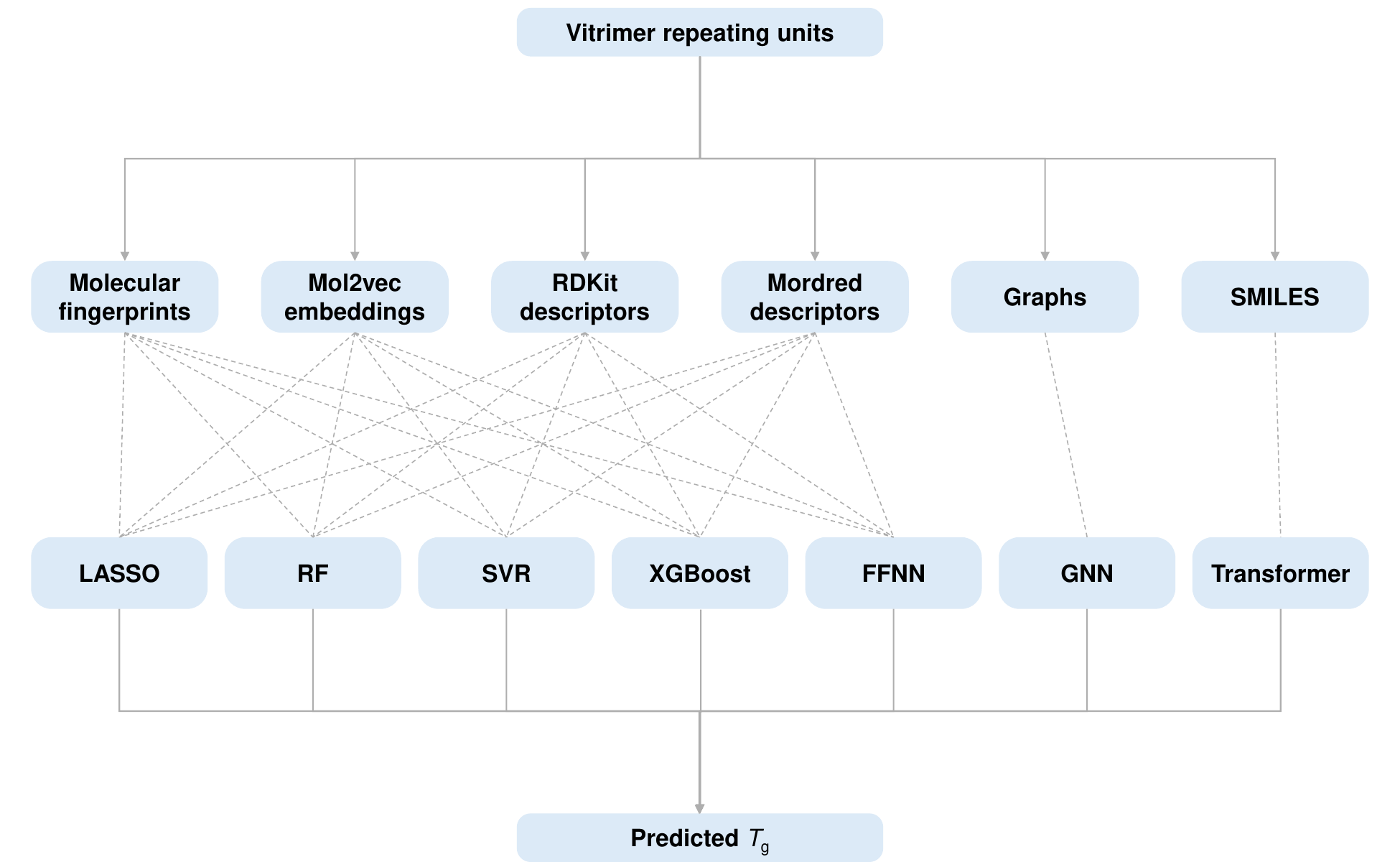}
\caption{Schematic overview of the 22 ML models covered in this work. Vitrimer repeating units are converted into six different types of feature representations, which are subsequently input into seven ML models for $T_\mathrm{g}$ prediction.}
\label{models}
\end{figure}

\begin{figure}[!ht]
\centering
\includegraphics[width=\linewidth]{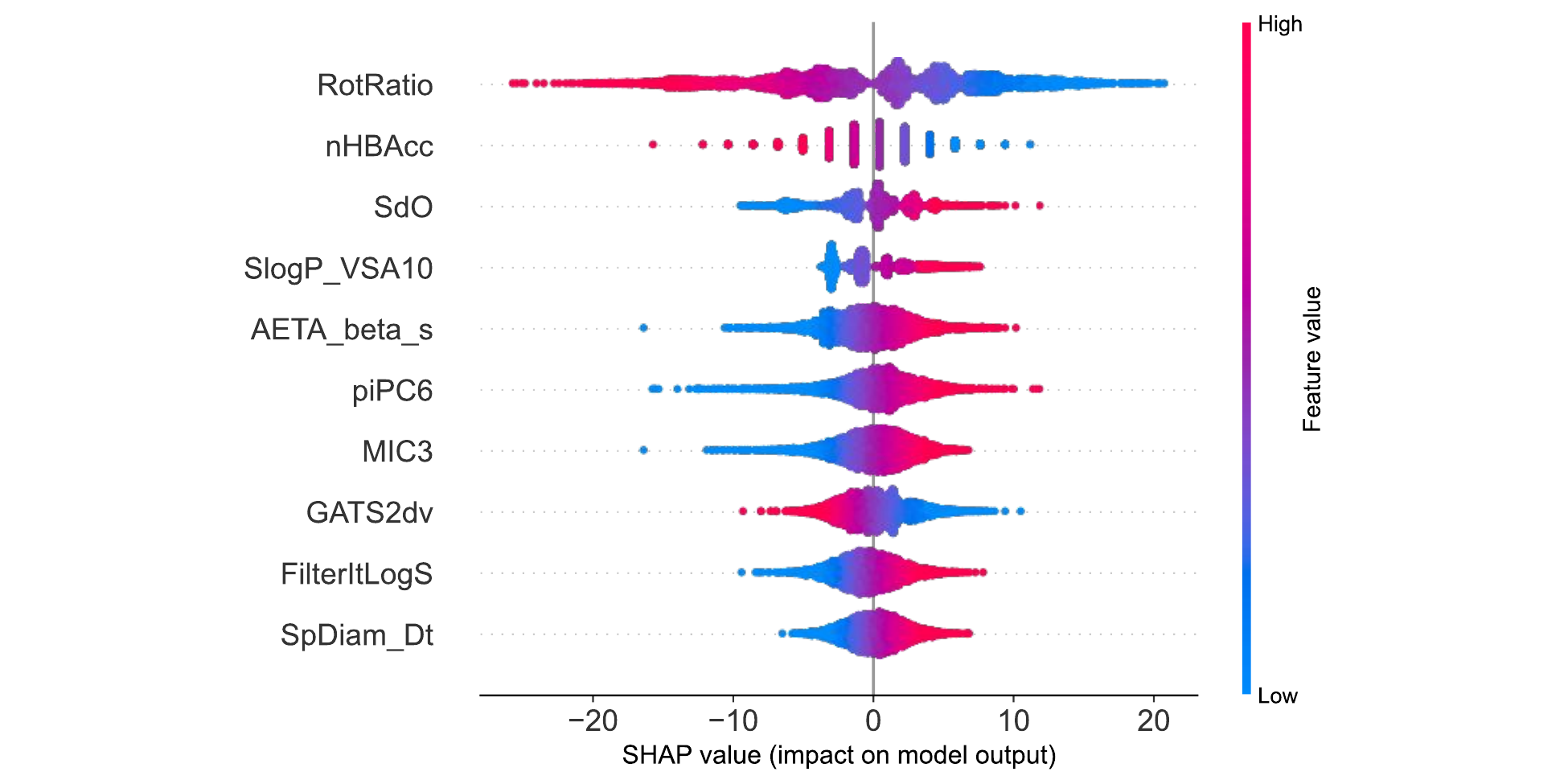}
\caption{SHAP values (feature importance scores) of ten most impactful Mordred descriptors.}
\label{shap_mordred}
\end{figure}

\begin{figure}[!ht]
\centering
\includegraphics[width=\linewidth]{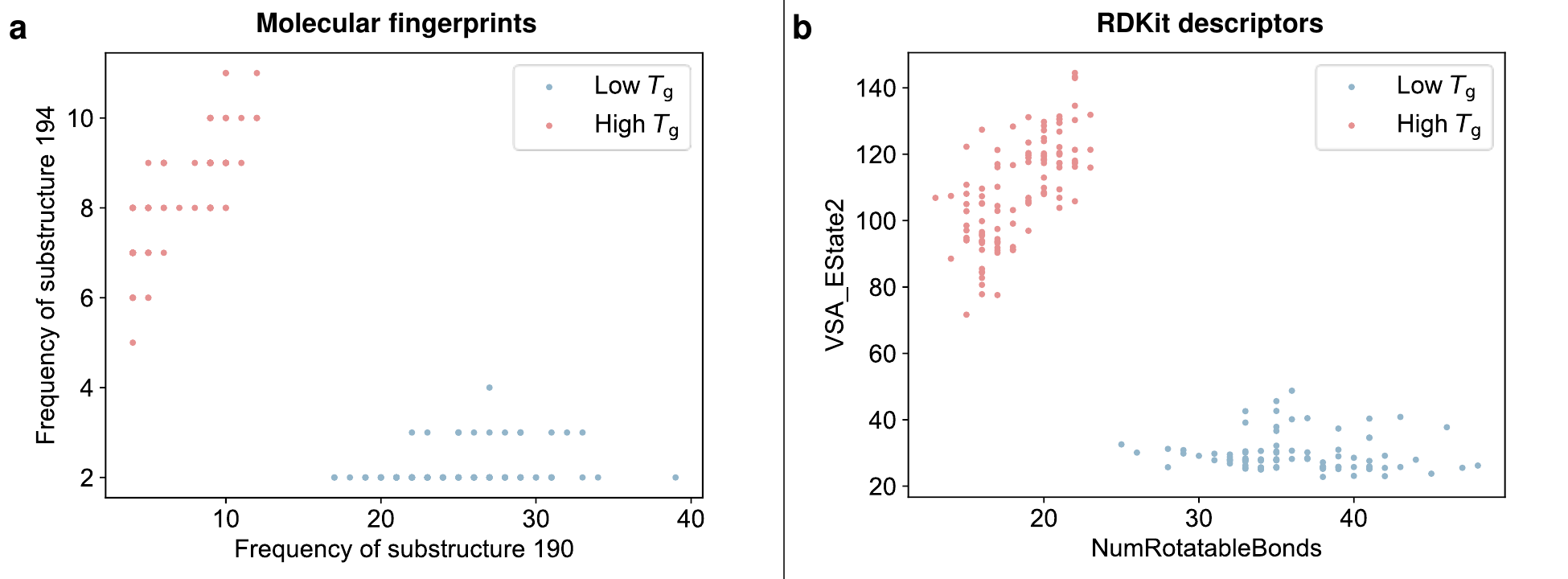}
\caption{Distribution of most impactful (a) substructures and (b) RDKit descriptors of discovered novel vitrimers.}
\label{screened_feature}
\end{figure}

\begin{figure}[!ht]
\centering
\includegraphics[width=\linewidth]{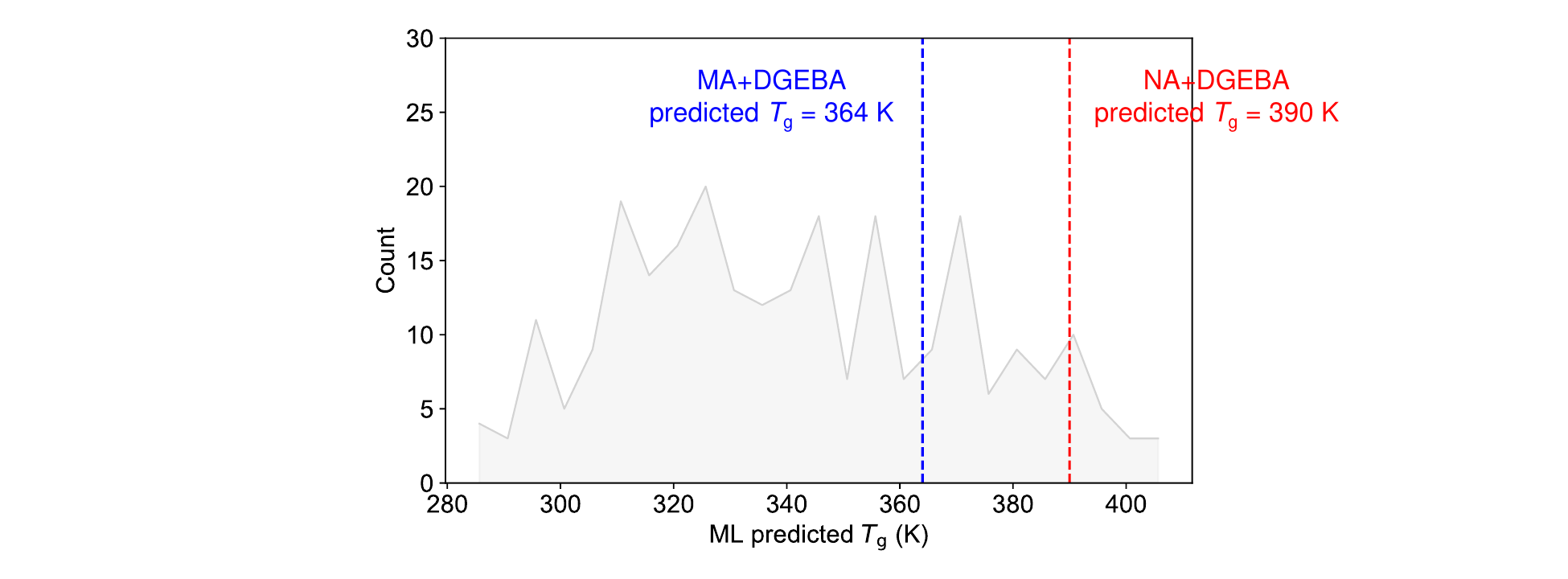}
\caption{Distribution of ML-predicted $T_\mathrm{g}$ of the unlabeled synthesizable dataset.}
\label{tg_synthesis}
\end{figure}

\clearpage

\bibliographystyle{elsarticle-num}
\bibliography{reference}